\documentstyle[preprint,aps,floats,epsfig,subeqn,amssymb]{revtex}

\newcommand{\be}{\begin{equation}}
\newcommand{\ee}{\end{equation}}
\newcommand{\bea}{\begin{eqnarray}}
\newcommand{\eea}{\end{eqnarray}}
\newcommand{\bml}{\begin{mathletters}}
\newcommand{\eml}{\end{mathletters}}

\begin{document}

\tighten

\preprint{DCPT-02/65}
\draft




\title{SU(2) gauged Skyrme-monopoles in scalar-tensor gravity}
\renewcommand{\thefootnote}{\fnsymbol{footnote}}
\author{ Yves Brihaye\footnote{Yves.Brihaye@umh.ac.be}}
\address{Facult\'e des Sciences, Universit\'e de Mons-Hainaut,
 B-7000 Mons, Belgium}
\author{Betti Hartmann\footnote{Betti.Hartmann@durham.ac.uk}}
\address{Department of Mathematical Sciences, University
of Durham, Durham DH1 3LE, U.K.}
\date{\today}
\setlength{\footnotesep}{0.5\footnotesep}

\maketitle

\begin{abstract}
Considering a Skyrme model with a peculiar gauging of the symmetry,
monopole-like solutions exist through a topological
lower bound. However, it was recently shown that these objects cannot form
bound states in the limit of vanishing Skyrme coupling.
Here we consider these monopoles in scalar-tensor gravity.
A numerical study of the equations
reveals that neither the coupling to gravity nor to the scalar dilaton
nor to dilaton-gravity leads to bound multimonopole states.
\end{abstract}
\medskip
\medskip
\newpage
\section{Introduction}
A few years ago topologically stable solutions with nonvanishing
magnetic flux were constructed \cite{bht} in a particular SU(2) gauged
SU(2) $\otimes$ SU(2) sigma model. This model differs essentially from
the (gauged) Skyrme model \cite{S} because in place of the usual
pion--mass potential of the latter, it is characterised by a potential
which leads to the breaking of the SU(2) symmetry down to U(1),
resulting in a monopole charge. The gauging prescription is given by
Eq. (6) of \cite{bht} and we employ the constrained non-linear sigma field 
$\phi^{\tilde{a}}$,
${\tilde{a}}=1,2,3,4$ with ($\sum_{\tilde{a}=1}^4 (\phi^{\tilde{a}})^2 = 1$) 
instead of $U=\phi^{\tilde{a}}\sigma_{\tilde{a}}\ 
,\ \ U^{\dagger}=\phi^{\tilde{a}}\tilde\sigma_{\tilde{a}}$.
The diagonal part of the SU(2) $\otimes$ SU(2) global symmetry is gauged
by means of the standard introduction of appropriate Yang-Mills fields (see \cite{bht} Eqs. (7)-(8) for details). 

The construction of the spherically symmetric solution
($n=1$) and of the axially symmetric solutions $(n > 1)$ \cite{bhtcorr} 
in the limit of vanishing Skyrme coupling 
indicates that the mass per winding number of the $n=2$ solution exceeds
that of the $n=1$ solution by roughly three percent,
leading to the conclusion that classical bound states are not possible.

On the other hand, in the much more popular bosonic part of the SU(2) Georgi-Glashow 
model which allows for 
monopole solutions with $n \geq 1$ \cite{hooft,rr}, 
it is well known that in the Prasad-Sommerfield limit of vanishing Higgs mass \cite{bps}
monopoles are non-interacting \cite{manton} and thus the  mass of the monopole
of topological charge $n$ is exactly $n$ times the mass of the single
monopole solution.
Recently, it was demonstrated that coupling
of the Georgi-Glashow model to gravity \cite{hkk}
and/or to a scalar field (the ``dilaton'') \cite{bh,bh2,h} results in bound multimonopole
solutions.

It thus seems sensible to couple the gauged non-linear
sigma model of \cite{bht} to gravity  and/or a dilaton
and to analyze if this new attracting interaction
can lead to bound states.

The numerical results we have obtained 
reveal that neither the coupling to gravity nor to the dilaton nor to dilaton-gravity
can render an attractive phase in the model studied here. We describe the model
in Section II. In Section III and IV, we describe the spherically symmetric, respectively
axially symmetric Ansatz and the obtained numerical results. We give our conclusions in Section V.

\section{The model}

The model is described by the Lagrangian of the field 
$\phi^{\tilde{a}}=(\phi^a,\phi^4)$, $a=1,2,3$ and an SU(2) valued
gauge field $A_{\mu}^a$~: 
\begin{equation}
\label{L}
{\cal L}_M=
-\frac{1}{4} G_{\mu \nu}^a G^{\mu \nu,a}  -{1\over 2}
\kappa_1^2 D_{\mu}\phi^{\tilde{a}} D^{\mu}\phi^{\tilde{a}} -{1\over2}\kappa_2^4
D_{[\mu}\phi^{\tilde{a}} D_{\nu]}\phi^{\tilde{b}} D^{[\mu}\phi^{\tilde{a}} D^{\nu]}\phi^{\tilde{b}} -V(\phi^4) \,
\end{equation}
with covariant derivative
\be
\label{cov}
D_{\mu} \phi^a =\partial_{\mu} \phi^a
+\varepsilon^{abc} A_{\mu}^b\phi^c \ , \ \ \
D_{\mu}\phi^{4} =\partial_{\mu} \phi^{4}\  \,
\ee
and field strength tensor
\begin{equation}
G_{\mu \nu}^a=\partial_{\mu} A_{\nu}^a - \partial_{\nu} A_{\mu}^a+
\varepsilon_{abc} A_{\mu}^b A_{\nu}^{c}
\end{equation}

In the following we work in the temporal gauge $A_0^a=0$ (thus the
solutions carry only a magnetic charge)
and assume $V = 0$. We are here premarily interested in the effect
of scalar-tensor gravity in the case of vanishing $\kappa_2$. The reason for this is that
for $\kappa_2\neq 0$, the results in \cite{bhtcorr} suggest
that the monopoles are already in an attractive phase in flat space and thus
the coupling to gravity would increase this attraction. We are here more interested
to see whether gravity and/or a dilaton can overcome the repulsion
of flat space (similar as it can in the case 
of non-vanishing Higgs self-coupling in the Georgi-Glashow model \cite{hkk}). 

The coupling of the matter field  to a (massless) dilaton field $\Psi$
consists in replacing ${\cal L}_M$ above by means of
\be
\label{LDIL}
{\cal L}_{M+D}=
-\frac{1}{4}  e^{2 \kappa \Psi} G_{\mu \nu}^a G^{\mu \nu,a} 
-\frac{1}{2} \partial_{\mu} \Psi \partial^{\mu} \Psi
-{1\over 2}\kappa_1^2 D_{\mu}\phi^{\tilde{a}}  D^{\mu}\phi^{\tilde{a}} \,
\ee
where the coupling constant to the dilaton is denoted by $\kappa$.

The coupling to gravity is done
by adding the Einstein-Hilbert action~:
\be
      S = S_G + S_M = \int {\cal L}_G \sqrt{- g} \ d^4 x
                     + \int{\cal L}_{M+D} \sqrt{-g} \ d^4 x
\ee
where ${\cal L}_{M+D}$  is defined above while 
\be
{\cal L}_G = \frac{1}{16 \pi G}  R
\ee
with $G$ denoting Newton's constant. Note that the dilaton can be decoupled by setting $\kappa=0$.

\section{Spherically symmetric solutions}
The spherically symmetric Ansatz for the field is described by

\be
\label{spha}
A_0^a=0 , \qquad
A_i^a=\frac{K(r)-1}{r}\ \varepsilon_{i a b}\ \hat x^{ b}\ ,
\ee
\be
\label{sphf}
\phi^{a}=\sin F(r)\ \hat x^{a}\ ,\qquad \phi^4=\cos F(r)\ .
\ee
for the matter fields,
\be
     ds^2 = -\sigma^2(r) N(r) dt^2 + N^{-1}(r) dr^2 + r^2 
(d \theta^2 + \sin^2 \theta d\varphi^2) \ , \ \ N(r)=1-\frac{2m(r)}{r}
\label{metric}
\ee
for the metric and $\Psi(\vec x)= \Psi(r)$ for the dilaton field. 

We use the following rescaling~:
\be
\label{rescale}
x \equiv r/\kappa_1 \ \ , \ \  
\mu(x)=m(r)/\kappa_1 \ \ , \ \
\Psi=\psi\kappa_1 \ .
\ee
The classical equations which are obtained after an algebra then
depend only on the dimensionless coupling constants $\alpha^2=
\sqrt{4 \pi G} \kappa_1$ and $\gamma=\kappa_1\kappa$:
\be
\mu' = \alpha^2(e^{2 \gamma \psi} N K'^2 + \frac{1}{2}N x^2 F'^2 
+ e^{2 \gamma \psi} \frac{(K^2-1)^2}{2 x^2} + (\sin F)^2 K^2 
+ \frac{1}{2} N x^2 (\psi')^2 )
\label{mueq}
\ee
\be
\sigma' = \alpha^2 x \sigma (e^{2 \gamma \psi}\frac{2 K'^2}{x^2} + F'^2  + (\psi')^2 )  
\label{sigmaeq}
\ee
\be
(e^{2 \gamma \psi} \sigma N K')' = \sigma K (e^{2 \gamma \psi} 
 \frac{K^2-1}{x^2} + (\sin F)^2) 
\label{keq}
\ee
\be
(x^2 \sigma N F')' = 2 \sigma K^2 \cos F \sin F
\label{feq}
\ee
\be
(x^2 \sigma N \psi')' = 2 \gamma \sigma e^{2 \gamma \psi} 
(N (K')^2 + \frac{(K^2-1)^2}{2 x^2})
\label{psieq}
\ee
These equations are solved subject to the following boundary 
conditions for the gravitating (resp. dilatonic) case :
\be
 K(0) = 1 \ \ , \ \ F(0) = 0 \ \ , \ \ K(\infty) = 0 \  \ , F(\infty) = \frac{\pi}{2}
\ee
\be
    \mu(0) = 0 \ \ , \ \ \sigma(\infty) = 1  \ \ , \ \ ({\rm resp.} \ \  \psi'(0)=0 \ \ , \psi(\infty)=0) 
\ee
Note that for  $\alpha=0$, $\gamma=0$  the equations admit a topological
soliton with classical mass $M\approx 1.475$ in units $4 \pi \kappa_1$.
The corresponding profiles of $K$ and $F$ are presented in Fig.~1.

If we set $\sin F(x) =: H(x)$, we recover the equations for the
gauge field function and dilaton function (\ref{keq}), (\ref{psieq}) of the SU(2) 
Einstein-Yang-Mills-Higgs-dilaton (EYMHD)
model \cite{bhk} in the BPS limit. The Einstein equations and the equation for the
scalar field function, however, are modified. Denoting by $P^{GG}_{\sigma,\mu,H}$
the rhs of the corresponding EYMHD equations (see \cite{bhk}) arising in the
Georgi-Glashow model, we rewrite (\ref{mueq}),
(\ref{sigmaeq}), (\ref{feq}) in terms of $H(x)$:
\begin{equation}
\mu'=P^{GG}_{\mu}+\frac{\alpha^2}{2}N x^2 \frac{H'^2 H^2}{1-H^2} \ , \ \ 
\sigma'=P^{GG}_{\sigma}+\alpha^2 x\sigma \frac{H'^2 H^2}{1-H^2}
\end{equation}
and
\begin{equation}
(x^2 \sigma N H')'=P^{GG}_{H}-\sigma\left(2 K^2 H^3 + x^2 N \frac{H
H'^2}{1-H^2}\right)
\end{equation}
with the abbreviations

\begin{equation}
P^{GG}_{\mu} = \alpha^2 \left(e^{2\gamma\psi}N(K')^2 + \frac{1}{2}N x^2(H')^2+
\frac{1}{2x^2}(K^2-1)^{2} e^{2\gamma\psi}+K^2 H^2 
+\frac{1}{2}Nx^{2}(\psi ')^2 \right)
\ , \label{dgl4} 
\end{equation}
\begin{equation}
P^{GG}_{\sigma}=\alpha^2 x \sigma \left(\frac{2(K')^2}{x^2}e^{2\gamma\psi}+(H')^2+(\psi ')^2
\right) \ , \ \ P^{GG}_{H}=2\sigma H K^2 \ .
\end{equation}

We first solved the equations in the gravitating case 
($\gamma = 0$)  for generic values of $\alpha$. The gauged skyrmion
solution of \cite{bht} is slightly deformed in a way very reminiscent
to the gravitating monopole \cite{bfm}~: the function $N$ 
develops a local minimum which becomes deeper with increasing $\alpha$.
For a maximal value of $\alpha$ this minimum is equal to zero at $x = x_m$,
the matter fields $K$ and $F$ vary mainly on $x\in [0,x_m]$ and
reach their asymptotic value for $x=x_m$. 
On the interval $x \in [x_m, \infty]$ the solution approaches an extremal 
Reissner-Nordstr\"om solution \cite{bfm}.
Our numerical analysis strongly suggests that $\alpha_{max} = x_m =1$,
$M(\alpha_{max}) = 1$. This maximal value is approached directly,
i.e. there is no backbending as in the gravitating GG-model 
for small value of the Higgs boson mass.

The value $\alpha_{max} = 1$ can be compared to those obtained in
the Georgi-Glashow (GG) model for gravitating monopoles.
It lies between the value of vanishing Higgs
boson mass (BPS limit) $\alpha^{0,GG}_{max} = 1.403$
and that of the infinite Higgs mass limit $\alpha^{\infty,GG}_{max}=1/\sqrt{2}$. 
We have not succeeded in finding an analytic
explanation for this phenomenon. In our case the equation for the $\phi$-field
remains non trivial while the corresponding equation
is just trivially fulfilled in the 
limit of infinite Higgs mass \cite{bfm} in the GG model and reproduces a power expansion of the solutions
which is more predictive.
 
The fact that the maximal value of $\alpha$ (i.e. $\alpha_{max} = 1.0$)
is much larger than in the GG-monopole case with infinite Higgs mass 
($\alpha^{\infty,GG}_{max}=1/\sqrt{2}$)
might be  understood by comparing the masses in both models for the
flat case. Defining $\beta \equiv M_H/M_W$ as the ratio of the Higgs boson mass 
and  the vector boson mass in the GG-model, it is well known \cite{bm} that
the monopole mass $M$ (in units of $4 \pi v/g^2$, $v$ being the vacuum expectation value of the Higgs field)
is such that
\be
            M(\beta = 0) = 1.0 \ \ \ , \ \ \ 
            M(\beta = 3) \approx 1.490 \ \ \ , \ \ \
            M(\beta = \infty) = 1.787 \ .
\ee
We note that in flat space the mass of our gauged Skyrme-monopole is very close to the mass
of the GG-monopole for $\beta \sim 3$.
On the other hand, it was shown \cite{bfm2} that $\alpha^{GG}_{max}(\beta=3)
\approx 1.0$ for the gravitating GG-monopoles. Since the mass density is the main parameter determining
the formation of a limiting black hole solution, this provides a consistent argument that 
$\alpha_{max}$ should be of the order of unity for the gravitating gauged Skyrme-monopole.

The profiles of the functions $K, F , N$ at the approach of the critical
value are shown in Fig. 1 together with the flat space profiles.
The mass of the gravitating solution as a function of
 $\sqrt{\alpha^2+\gamma^2}|_{\gamma=0}=\alpha$
is plotted in Fig. 2 ($n=1$, circles).

We then solved the equations in the purely dilatonic case ($\gamma \neq 0$,
$\alpha = 0$) which implies $N=A\equiv 1$. We found a similar pattern as in the Georgi-Glashow
model coupled to a dilaton \cite{forgacs}~: the flat solution is gradually deformed by the dilaton
field, the value $\psi(0)$ is negative and decreases while $\gamma$ increases
and becomes infinite in the limit $\gamma \rightarrow \gamma_{max}$.
Our numerical analysis strongly suggest that $\gamma_{max}=1$.
The mass of the dilatonic solution is shown in Fig. 2 as  function of 
$\sqrt{\alpha^2+\gamma^2}|_{\alpha=0}=\gamma$
($n=1$, triangles). It hardly differs from its counterpart
for gravitating solutions as function of $\sqrt{\alpha^2+\gamma^2}|_{\gamma=0}=\alpha$.
The similarities between the influence
of a gravitating field and of a dilatonic field on solitons was noticed previously
in \cite{forgacs}  and in \cite{bh} for SU(2) Yang-Mills-Higgs theories.
Our results demonstrate that this analogy persists for
Skyrme-like solutions. It is furthermore easy to see
that the  correspondence between the $tt$-component of the metric and
the dilaton found previously for the GG-monopoles \cite{bhk} is also valid here. 

More remarkably, solving the equations for fixed $\eta \equiv \alpha/ \gamma$ leads 
to identical curves for the mass $M(\sqrt{\alpha^2 + \gamma^2}$) (within the numerical inaccuracy) 
irrespective of of the choice of $\alpha\neq 0$ and $\gamma\neq 0$. Thus plotting the
mass as function of $\sqrt{\alpha^2 + \gamma^2}$ leads to a  curve very similar
to the pure gravitating and pure dilatonic one. This is demonstrated in Fig. 2 ($n=1$, hexagons).
In the limit of critical coupling, the solutions for $\alpha\neq 0$, $\gamma\neq 0$ bifurcate with the
Einstein-Maxwell-dilaton (EMD) solutions, very similar to the solutions in SU(2) 
Einstein-Yang-Mills-Higgs-dilaton (EYMHD) theory \cite{bhk}.

\section{Axially symmetric solutions}
The axially symmetric Ansatz for the metric in isotropic coordinates reads:
\begin{equation}
ds^2=
  - f dt^2 +  \frac{m}{f} \left( d \tilde{r}^2+ \tilde{r}^2d\theta^2 \right)
           +  \frac{l}{f} \tilde{r}^2\sin^2\theta d\varphi^2
\ ,  \end{equation}
The functions  $f$, $m$ and $l$ now depend on $\tilde{r}$ and $\theta$. 
If $l=m$ and $f$ only depend on $\tilde{r}$, this metric reduces to the
spherically symmetric metric in isotropic coordinates and comparison
with the metric in (\ref{metric}) yields the coordinate transformation 
\cite{hkk1}:
\begin{equation}
\frac{d\tilde{r}}{\tilde{r}}=\frac{1}{\sqrt{N(r)}}\frac{dr}{r}
\end{equation}
For the gauge fields we choose the purely magnetic Ansatz:
\begin{equation}
{A_t}^a=0 \ , \ \ \  {A_{\tilde{r}}}^a=\frac{H_1}{\tilde{r}}{v_{\varphi}}^a 
\ , \end{equation}
\begin{equation}
{A_{\theta}}^a= (1-H_2) {v_{\varphi}}^a
\ , \ \ \ \ 
{A_{\varphi}}^a=-n\sin\theta \left(H_3{v_{\tilde{r}}}^a+
(1-H_4) {v_{\theta}}^a \right)
\ . \end{equation}
while for the sigma-field, the Ansatz reads \cite{bhtcorr}
\begin{equation}
{\phi}^{a}= (\phi_1 {v_{\tilde{r}}}^{a}+\phi_2 {v_{\theta}}^{a}) \ , \ \ \phi^4=\sqrt{1-(\phi_1^2+\phi_2^2)}
\ . \end{equation}
The vectors $\vec{v}_{\tilde{r}}$,$\vec{v}_{\theta}$ and $\vec{v}_{\varphi}$
are given by:
\begin{eqnarray}
\vec{v}_{\tilde{r}}      &=& 
(\sin \theta \cos n \varphi, \sin \theta \sin n \varphi, \cos \theta)
\ , \nonumber \\
\vec{v}_{\theta} &=& 
(\cos \theta \cos n \varphi, \cos \theta \sin n \varphi,-\sin \theta)
\ , \nonumber \\
\vec{v}_{\varphi}   &=& (-\sin n \varphi, \cos n \varphi,0) 
\ .\label{rtp} \end{eqnarray} 
The dilaton field $\Psi$ now depends on $\tilde{r}$ and $\theta$ \cite{bh}:
\begin{equation}
\Psi=\Psi(\tilde{r},\theta)
\end{equation}
At the origin, the boundary conditions read:
\begin{equation}
\partial_{\tilde{r}}f(0,\theta)=\partial_{\tilde{r}}l(0,\theta)=
\partial_{\tilde{r}}m(0,\theta)=0,\ \ \partial_{\tilde{r}}\Psi(0,\theta)=0
\end{equation}
\begin{equation}
H_i(0,\theta)=0,\ i=1,3 ,\ \ H_i(0,\theta)=1,\ i=2,4,\ \
\phi_i(0,\theta)=0,\ i=1,2 
\end{equation}
At infinity, the requirement for finite energy and asymptotically
flat solutions leads to the boundary conditions:
\begin{equation}
f(\infty,\theta)=l(\infty,\theta)=m(\infty,\theta)=1,\ \ \Psi(\infty,\theta)=0
\end{equation}
\begin{equation}
H_i(\infty,\theta)=0,\ i=1,2,3,4 ,\ \ \phi_1(\infty,\theta)=1,\ \
\phi_2(\infty,\theta)=0 
\end{equation}
In addition,
boundary conditions on the symmetry axes (the $\rho$- and
$z$-axes) have to be fulfilled.
On both axes:
\begin{equation} 
H_1=H_3=\phi_2=0
\end{equation}
and
\begin{equation}
\partial_\theta f=\partial_\theta m=\partial_\theta l 
=\partial_\theta H_2=\partial_\theta H_4=
\partial_\theta \phi_1=\partial_\theta \Psi=0
\end{equation}

After a similar rescaling as in (\ref{rescale}) we have solved  
the classical equations numerically for
$n=1,2$. The energy per winding number $E/n$ of the solutions
is presented in Fig. 2 for the dilatonic (triangles),  
gravitating (circles) and dilatonic-gravitating (hexagons) case as function of
$\sqrt{\alpha^2+\gamma^2}$.
Like for $n=1$, the similarity of the dependence of the masses 
on the parameter $\sqrt{\alpha^2 + \gamma^2}$ is striking.

The other main point of this figure is that
we find  $\Delta \equiv E_2/2 - E_1 > 0$ for all three cases.
Although the difference $\Delta$ is maximal in the flat case
($\gamma=0$ and/or $\alpha=0$) (it is about three percent of the mass
of the classical lump) it decreases, as expected, for $\alpha >0$ and/or $\gamma >0$. 
However,  neither the gravitating nor the dilatonic nor the dilatonic-gravitating
interaction is
strong enough to overcome the repulsion.

Let us finally remark, that our numerical results strongly suggest (although with less
accurancy than in the spherically symmetric case $n=1$) that the $n=2$ solution for $\gamma=0$
bifurcates into an extremal Reissner-Nordstr\"om  black hole at exactly
$\alpha_{max} = 1$. The plot of the quantity $f(0,\theta)$ as a function
of $\alpha$  clearly shows that it tends to zero  
in the limit $\alpha\rightarrow 1$ and that at the same time the angle dependence of $f(r,\theta)$
vanishes.
Since the extremal Reissner-Nordstr\"om solution in isotropic coordinates
has a horizon located at $x=0$, the
pattern described above is a signature of a bifurcation into an 
extremal RN black hole \cite{hkk,hkk1}.
Similarly, the corresponding extremal solutions in isotropic coordinates are reached
in the limit of critical coupling in the pure dilatonic, respectively dilatonic-gravitating case.

\section{Concluding remarks}
The gauged skyrmion model proposed in \cite{bht}
and the corresponding topological solutions constitute
an alternative to the celebrated Georgi-Glashow
model and its (multi-)monopoles.  
It supports a magnetic charge and, because of a topological
inequality,  exists even in absence of a Skyrme term ($\kappa_2=0$).

In this paper, we were mainly interested in the questions whether bound
states of gauged Skyrme-monopoles are possible. 
In the absence  of a Skyrme term, bound states are not possible in flat
space \cite{bhtcorr}. A natural step then was to study  
whether gravitating or dilatonic
(or both) interactions can lead to an attractive phase (similarly as in the
Georgi-Glashow model for small enough Higgs boson mass).
Unfortunately, we found  that this is not possible and likely, the construction
of bound states of gauged Skyrme-monopoles will require a non-vanishing
Skyrme term (for which in flat space there already seems
to be an attraction).

However, we believe that the results presented here  reveal a potentially interesting property
of the model: it seems that the limit $\alpha \rightarrow 1.0$
corresponds to a bifurcation of the $n=1$ and $n=2$ solutions
into a Reissner-Nordstr\"om black hole with magnetic charge $n$ and
mass $n$ (in suitable units). Attempts to explain 
the occurence of the value $\alpha_{max}=1$ algebraically are under investigation.

\bigskip
\noindent
{\bf Acknowledgements:} We gratefully acknowledge discussions
with D. H. Tchrakian. B. H. was supported by the EPSRC.

\small{

\begin{figure}
\centering
\epsfysize=15cm
\mbox{\epsffile{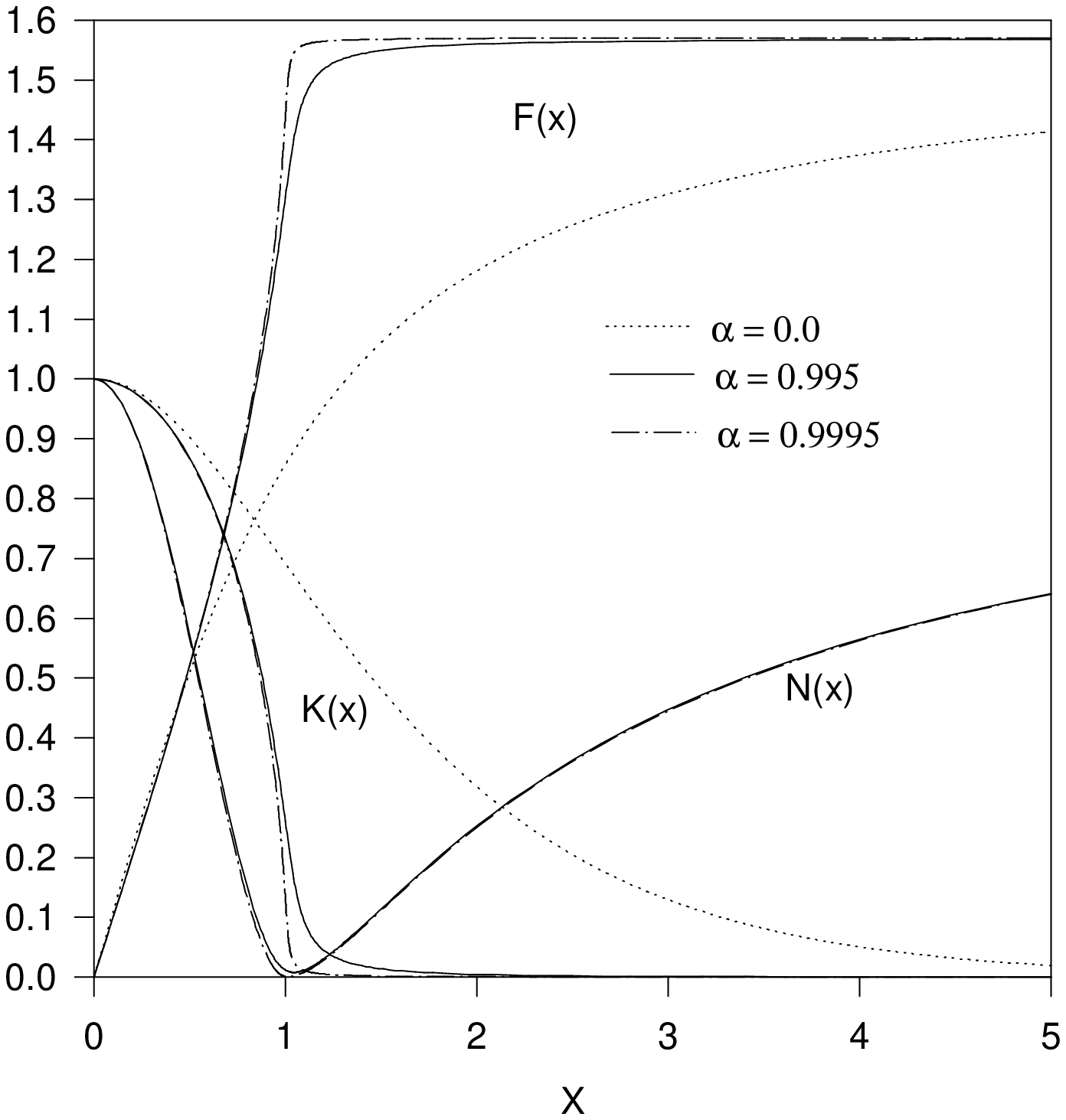}}
\caption{The gauged Skyrme-monopole functions  $K(x)$, $F(x)$ and
$N(x)$ are shown as functions of the dimensionless coordinate $x$
for $\alpha = 0.995$, $\gamma=0$ (solid) and for
$\alpha=\alpha_{max}\approx 0.9995$, $\gamma=0$ (dotted-solid). 
For comparison, the corresponding flat space ($\alpha=0$, $\gamma=0$) functions
are also shown (dotted).}
\end{figure}
\newpage
\begin{figure}
\centering
\epsfysize=15cm
\mbox{\epsffile{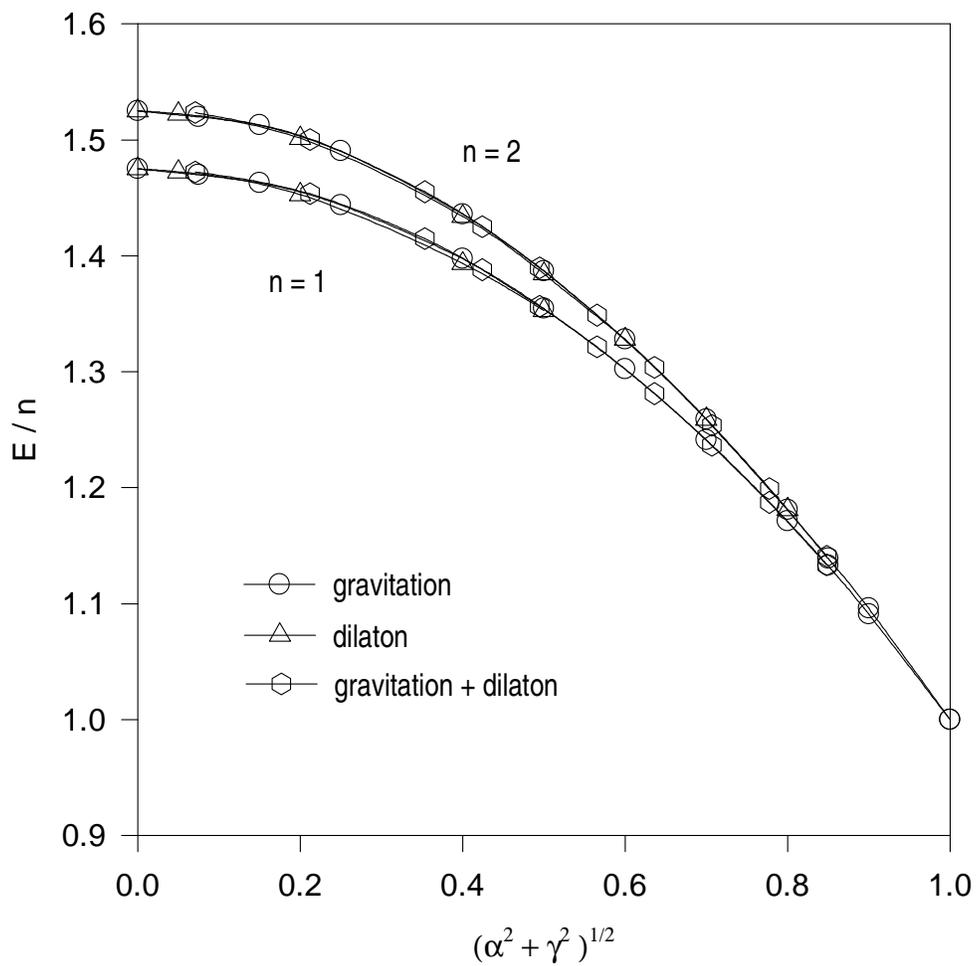}}
\caption{The energy per winding number $E/n$ 
of the gravitating 
solution, $\alpha\neq 0$, $\gamma=0$, (circles),
the dilatonic solution, $\alpha=0$, $\gamma \neq 0$, (triangles),  and the
dilatonic-gravitating solution (hexagons), respectively, is presented as 
function of $(\alpha^2 + \gamma^2)^{1/2}$ for $n=1, 2$.}
\end{figure}

\end{document}